\documentclass{ws-ijmpd}

\begin{document}

\markboth{Hossein, Rahaman, Naskar, Kalam and Ray}
{Anisotropic Compact stars}

%
\catchline{}{}{}{}{}
%

\title {Anisotropic Compact stars with variable cosmological constant}

\author{Sk. Monowar Hossein}
\address{Department of
Mathemaics, Aliah University, Sector - V , Salt Lake,  Kolkata -
700091, India\\sami\_milu@yahoo.co.u}

\author{Farook Rahaman}
\address{Department of
Mathematics, Jadavpur University, Kolkata 700 032, West Bengal,
India\\farook@iucaa.ernet.in}

\author{Jayanta Naskar}
\address{Department of Physics,
Purulia Zilla School,Purulia, West Bengal, India\\jayanta86@gmail.com}

\author{Mehedi Kalam}
\address{Department of
Physics, Aliah University, Sector - V, Salt Lake,  Kolkata -
700091, India\\mehedikalam@yahoo.co.in}

\author{Saibal Ray}
\address{Department of Physics,
Government College of Engineering \& Ceramic Thechnology, Kolkata
700 010, West Bengal, India\\saibal@iucaa.ernet.in}

\maketitle

\begin{history}
\received{Day Month Year}
\revised{Day Month Year}
\comby{Managing Editor}
\end{history}

\begin{abstract}
Recently the small value of the cosmological constant and its
ability to accelerate the expansion of the Universe is of great
interest. We discuss the possibility of forming of anisotropic
compact stars from this cosmological constant as one of the
competent candidates of {\it dark energy}. For this purpose we
consider the analytical solution of Krori and Barua metric. We
take the radial dependence of cosmological constant and check all
the regularity conditions, TOV equations, stability and surface
redshift of the compact stars. It has been shown as conclusion
that this model is valid for any compact star and we have cited
$4U~1820-30$ as a specific example of that kind of star.

\end{abstract}

\keywords{General Relativity; cosmological constant; compact star}

\section{Introduction}
The study of general relativistic compact objects is of great
interest for a long time. The theoretical investigation of superdense
stars has been done by several authors using both analytical and
numerical methods. \cite{Bowers1974} emphasized on the importance
of locally anisotropic equations of state for relativistic fluid
spheres for generalizing the equations of hydrostatic equilibrium
to include the effects of local anisotropy. They showed that
anisotropy may have non-negligible
effects on such parameters like maximum equilibrium mass and surface
redshift. \cite{Ruderman1972} has investigated the stellar
models and argued that the nuclear matter may have anisotropic features at
least in certain very high density ranges ($> ~ 10^{15}$gm/c.c.),
where the nuclear interaction must be treated relativistically.

Einstein introduced the cosmological constant into his field
equations to make them consistent with Mach's principle and to
get a non-expanding static solutions of the Universe. If matter
is the source of inertia, then in absence of it there should not
be any inertia. In view of that he introduced the cosmological
constant $\Lambda$ to eliminate inertia when matter is
absent \cite{Einstein1917}. With the successes of FRW cosmology
and later on experimental verification of expanding Universe by Hubble
the idea of $\Lambda$ was abandoned by Einstein himself.

However, in modern cosmology, recent measurement conducted by WMAP
indicates that $73\%$ of the total mass-energy of the Universe is
{\it dark energy} \cite{Perlmutter1998,Riess2004}. Dark energy
theory is therefore thought to be the most acceptable one to
explain the present phase of the accelerated expansion of the
Universe. The erstwhile Cosmological constant adopted by Einstein
is a good candidate to explain this dark energy in the
cosmological realm. However, even in the astrophysical point of
view we would like to go back to the idea of \cite{Einstein1917}
that $\Lambda$ has something common in matter distribution.

As a start of the astrophysical objects of compact nature
(either as neutron stars or strange stars or others) let us mention some
of the important works done by \cite{Dymnikova2002},
\cite{Egeland2007} and \cite{Burdyuzha2009} where $\Lambda$ have taken
as a source of matter distribution of the stars. \cite{MaK2000}
have calculated the mass-radius ratio for compact relativistic star
in the presence of cosmological constant. On the other hand,
anisotropy have been introduced as key feature in the configuration of
the compact stars which enabled several investigators to model the objects
physically more viable \cite{Varela2010,Rahaman2010,Rahaman2011,Rahaman2012,Mehedi2012}.
In the studies of compact stars some other notatble works
with different aspects are as follows: \cite{Chodos1974},
\cite{Li1999a,Li1999b}, \cite{Mak2002,Mak2003}, \cite{Usov2004}.

Motivating with the above investigations we take the cosmological
constant as a source of matter and study the structure of compact stars
and arrived at the conclusion that incorporation of $\Lambda$ describes
the well known compact stars, for examples neutron stars, white dwarf stars,
X ray buster 4U 1820-30, Millisecond pulsar SAX J 1808.4-3658 etc.,
in a good manner. In this article, therefore, we have proposed a model for
anisotropic compact stars which satisfies all the energy conditions,
TOV-equations and other physical requirements. We also investigate the
stability, mass-radius relation and surface redshifts for our model and
found that their behaviour is well matched with the compact stars.

\section{Non-singular Model for Anisotropic Compact Stars}

The general line element for a static spherically symmetric space-time in
 \cite{Krori1975} model (henceforth KB) is given by
\begin{equation}
ds^2=-e^{\nu(r)}dt^2 + e^{\lambda(r)}dr^2 +r^2
(d\theta^2 +sin^2\theta d\phi^2), \label{eq1}
\end{equation}
with $\lambda(r)=Ar^2$ and $\nu(r) = Br^2 + C$ where $A$, $B$ and $C$ are
 arbitrary constants to be determined on  physical grounds.
 The interior of the anisotropic compact star may be expressed in the
standard form as
\[T_{ij}=diag(\rho,-p_r,-p_t,-p_t),\]
where $\rho$, $p_r$ and $p_t$ correspond to the energy density,
normal pressure and transverse pressure respectively.

We take the cosmological constant as radial dependence i.e. $\Lambda ~=~
\Lambda(r) ~=~\Lambda_r$ (say). Therefore, the Einstein field equations
for the metric (\ref{eq1}) are obtained as (where natural units $G = c=1$)
\begin{eqnarray}
\label{eq2}
 8\pi\rho+ \Lambda_r &=&
e^{-\lambda}\left(\frac{\lambda^\prime}{r}-\frac{1}{r^2}\right) +
\frac{1}{r^2},\\
\label{eq3}
8\pi p_r - \Lambda_r &=&
e^{-\lambda}\left(\frac{\nu^\prime}{r}+\frac{1}{r^2}\right) -
\frac{1}{r^2},\\
\label{eq4}
8\pi\ p_t - \Lambda_r &=&
\frac{e^{-\lambda}}{2}\left[\frac{{\nu^\prime}^2 -
\lambda^{\prime}\nu^{\prime}}{2}
+\frac{\nu^\prime-\lambda^\prime}{r}+\nu^{\prime\prime}\right].
\end{eqnarray}

To get the physically acceptable stellar models, we propose that the
radial pressure of the compact star is proportional to the matter density i.e.
\begin{equation}
\label{eq5}
p_r = m\rho,~~~ m>0,
\end{equation}
 where $m$ is the equation of state parameter.

Now, from the metric (\ref{eq1}) and equations (\ref{eq2}) - (\ref{eq5}),
we get the energy density $\rho$, normal pressure $p_{r}$, tangential pressure
$p_{t}$ and cosmological parameter $\Lambda_{r}$, respectively as
 \begin{eqnarray}
\rho &=& \frac{2(A+B)}{8\pi (m+1)}e^{-Ar^2} >0, \label{eq6}\\
p_{r} &=& \frac{m(A+B)}{4\pi (m+1)}e^{-Ar^2} >0, \label{eq7}\\
p_{t} &=& \frac{1}{8\pi (m+1)}[e^{-Ar^2}((m+1)(B^2
-AB)r^2 \nonumber \\
& & +m(2B+A)- A - \frac{(m+1)}{r^2})+\frac{(m+1)}{r^2}]. \label{eq8}\\
\Lambda_{r} &=& \frac{e^{-Ar^2}}{m+1}\left[2(Am-B)-\frac{m+1}{r^2}\right]+\frac{1}{r^2}.
\end{eqnarray}

Also, the equation of state (EOS) parameters corresponding to normal and
transverse directions can be written as
\begin{equation}
\label{eq11} \omega_r(r)   =  m,
\end{equation}
\begin{eqnarray}
\label{eq12} \omega_t(r) =
\frac{1}{2(A+B)}\left[\frac{e^{Ar^2}(m+1)}{r^2} -
\frac{(m+1)}{r^2} \right.\nonumber\\
\left.+ (m+1)(B^2-AB)r^2+m(2B+A-\frac{A}{m})\right].
\end{eqnarray}

\section{Physical Analysis}

In this section we will discuss the following features of our
model:

\subsection{Anisotropic Behavior}

From the equations (\ref{eq6}) and (\ref{eq7}) we have
\[ \frac{d\rho}{dr}= -\left[\frac{A(A+B)}{2\pi (m+1)}re^{-Ar^2}\right] < 0,\]
~~~and~~~
\[\frac{dp_r}{dr} < 0.\]

The above results and Figs. (1) and (2) conclude that, at $r=0$,
our model provides
\[ \frac{d\rho}{dr}=0,~~ \frac{dp_r}{dr} = 0, \]
\[ \frac{d^2 \rho}{dr^2}=-\frac{A(A+B)}{2\pi (m+1)} < 0,\]
~~~and~~~
\[\frac{d^2 p_r}{dr^2} = -\frac{mA(A+B)}{2\pi (m+1)} < 0.\]
which indicate maximality of central density and central pressure.

\begin{figure}
\centering
\includegraphics[scale=.3]{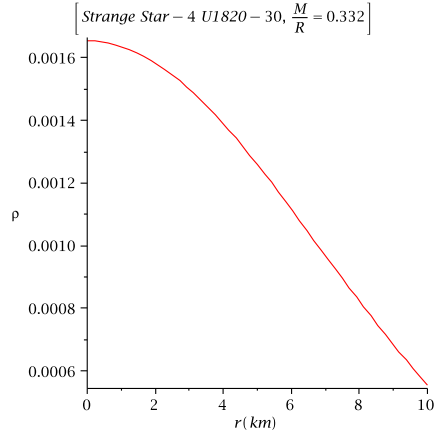}
\caption{Density variation at the stellar interior of Strange star candidate $4U~1820-30$}
\label{fig:1}
\end{figure}

\begin{figure}
\centering
\includegraphics[scale=.3]{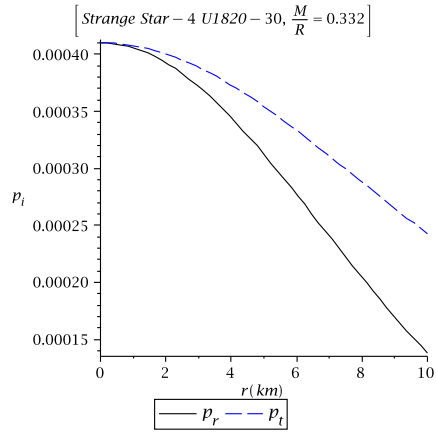}
\caption{Radial and Transverse Pressure variation at the
stellar interior of Strange star candidate $4U~1820-30$}
    \label{fig:2}
\end{figure}

It is interesting to note that, like normal matter distribution,
the bound on the effective EOS in this construction is given by
$0< \omega_i(r)<1$, (see Fig. (3)) despite of the fact that star is
constituted by the combination of ordinary matter and effect of
$\Lambda$. Here figures are drawn taking m=$\frac{1}{3}$.

\begin{figure}
\centering
\includegraphics[scale=.3]{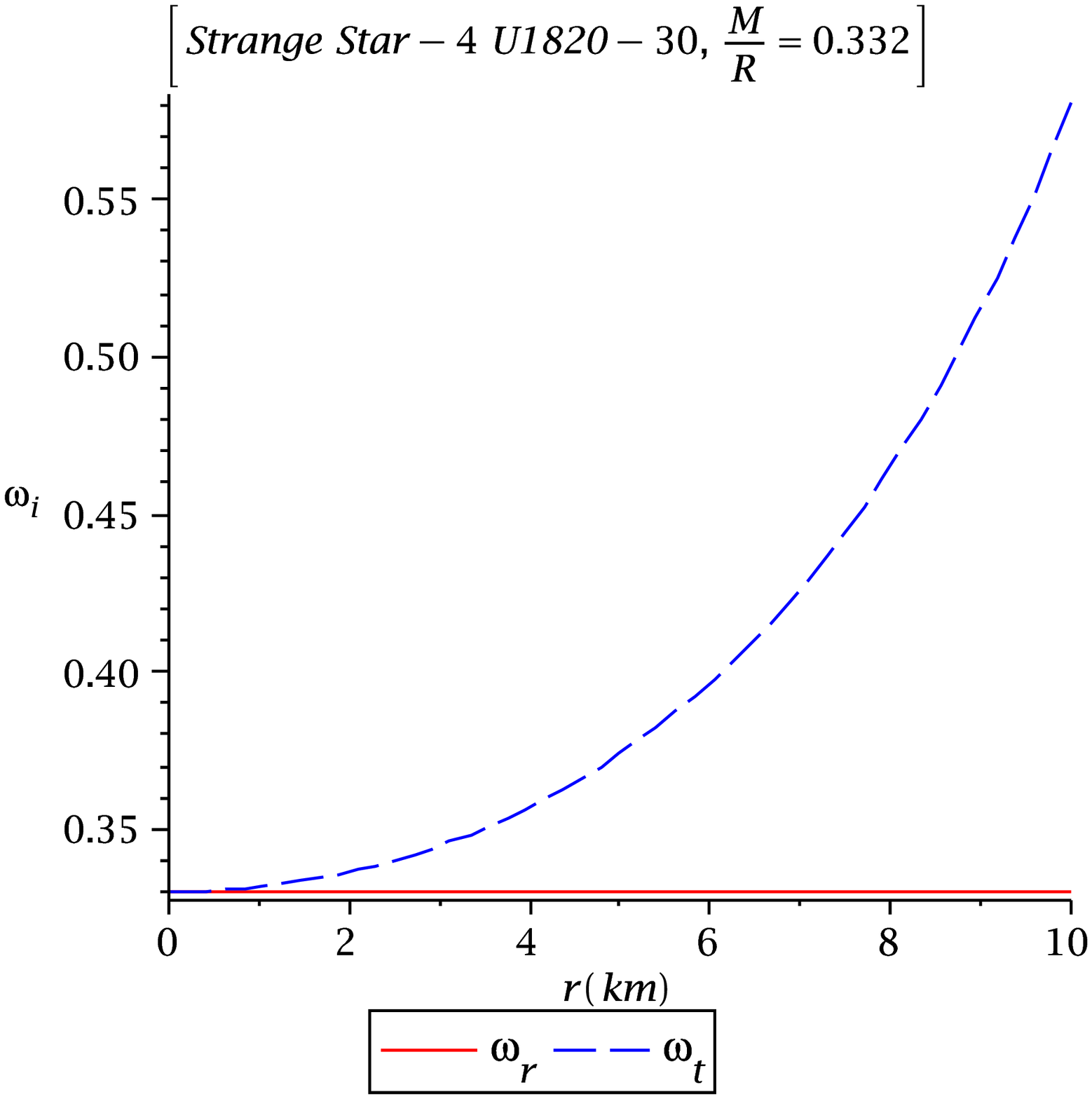}
\caption{ Variation of equation of state parameter with distance
of Strange star candidate $4U~1820-30$}
\label{fig:3}
\end{figure}

The measure of anisotropy, $\Delta  =
\frac{2}{r}\left(p_{t}-p_{r}\right)$, in this model is obtained as
\begin{eqnarray}
\label{eq13} \Delta  =
 \frac{1}{4\pi r
}\left[e^{-Ar^2}\left((B^2 -AB)r^2  -A-\frac{1}{r^2}\right)+\frac{1}{r^2}\right].
\end{eqnarray}
The `anisotropy' will be directed outward when $P_t > P_r$ i.e. $\Delta
> 0 $, and inward when $P_t < P_r$ i.e. $\Delta < 0$.
It is apparent from the Fig. (4) of our
model that a repulsive `anisotropic' force ($\Delta > 0$) allows the
construction of more massive distributions.

\begin{figure}
\centering
\includegraphics[scale=.3]{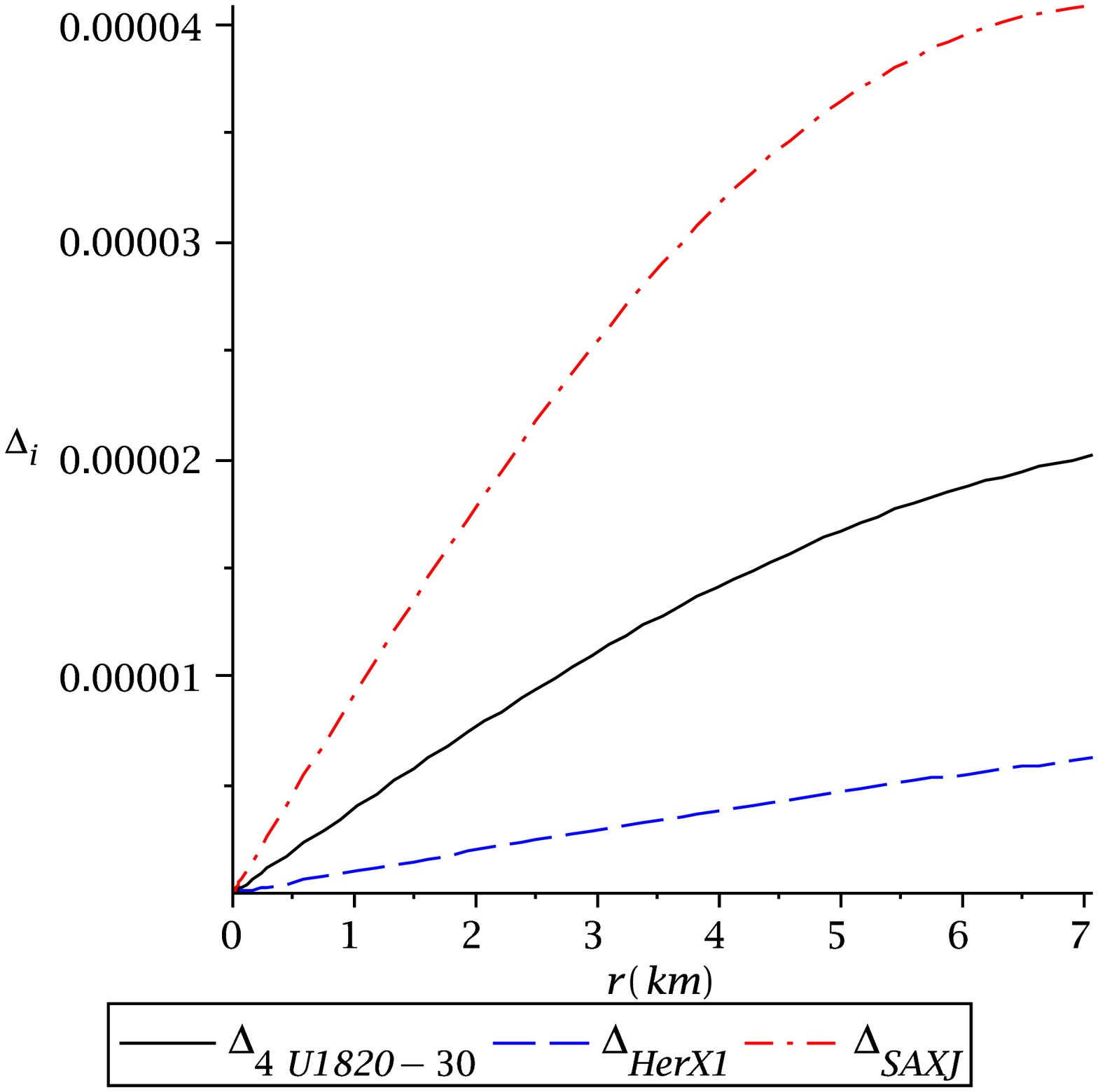}
\caption{ Comparison of anisotropic behaviours at the stellar interior of
Strange star candidates $4U~1820-30$, $Her~X-1$ and  $SAX~J~1808.4-3658$}
\label{fig:4}
\end{figure}

\subsection{Matching Conditions}
Here we match the interior metric to the Schwarzschild exterior
\begin{eqnarray}
ds^{2}=-\left(1 - \frac{2M}{r} \right)dt^2
+ \left(1 - \frac{2M}{r} \right)^{-1}dr^2 \nonumber \\+ r^2(d\theta^2+\sin^2\theta d\phi^2),
\label{eq20}
\end{eqnarray}
at the boundary $r= b$ (radius of the star). Assuming the boundary conditions
$p_r(r=b) = 0$ and $\rho( r=0) = \rho_{0}$, where $\rho_{0}$ is the
central density, we have
\begin{eqnarray}
A &=& \frac{8 \pi \rho_{0}}{3},\label{eq18} \\
B &=& \frac{1}{2b^2} \left[ e^{\frac{8 \pi \rho_{0}}{3}
b^2}-1\right].\label{eq19}
\end{eqnarray}

\begin{table*}
\centering
\begin{minipage}{140mm}
\caption{Values of the model parameters for
different Strange stars.}\label{tbl}
\begin{tabular}{@{}lrrrrrr@{}}
\hline
Strange Quark Star & $M$($M_{\odot}$) & $R$(km) & $\frac{M}{R}$ & $A$ (km$^{-2}$) & $B$ (km$^{-2}$) & $\rho_{0}$(km$^{-2}$) \\ \hline
Her X-1 & 0.88 & 7.7 & 0.168 & 0.006906276428 & 0.004267364618 & 0.0008247941592  \\
SAX J 1808.4-3658(SS1) & 1.435 & 7.07 & 0.299 & 0.01823156974 & 0.01488011569 & 0.002177337150 \\
4U 1820-30 & 2.25 & 10.0 & 0.332 & 0.01090644119 & 0.009880952381 & 0.001302520843\\
\hline
\end{tabular}
\end{minipage}
\end{table*}

At this juncture, let us first evaluate some reasonable set for
values of $A$ and $B$.
According to \cite{Buchdahl1959}, the maximum allowable
compactness (mass-radius ratio) for a fluid sphere is given by
$\frac{2M}{R} <\frac{8}{9}$. Values of the model parameters for different
Strange stars (Table 1) show the acceptability of our model.

We have also verified with different sets of mass and radius
leading to solutions for the unknown parameters which satisfy the
following conditions through out the configuration:
\[\rho \geq 0, ~~~\rho+p_r \geq 0, ~~~\rho+p_t \geq 0,~~~\rho+p_r+2p_t \geq 0,\]
$\rho > |p_r|$ and $\rho > |p_t|$.

 Null Energy Condition (NEC), Weak Energy Condition (WEC), Strong Energy
Condition (SEC) and Dominant Energy Condition (DEC) i.e. all the
energy conditions for our particular choices of the values of mass
and radius are satisfied.

The anisotropy, as expected, vanishes at the centre, i.e. $r=0$, $p_t =
p_r = p_0=\frac{m(A+B)}{4\pi (m+1)}$. The energy density
and the two pressures are also well behaved in the interior of
the stellar configuration.

\subsection{Matter Density and Pressure}
From the equations (6),  (14) and (15)  we have
\[\rho = \rho_{0}e^{(-\frac{8\pi \rho_{0}}{3})r^2}.\]

Since it is well known that $0<m<1$ and $\frac{M}{R}~<~\frac{4}{9} $
for a compact star with radius $R$ and mass $M$, hence our model
 suggests a range for matter density of the star, which is
\[e^{-Ab^2}\frac{A+B}{8\pi}~<\rho~<\frac{3}{\pi b^2}.\]

According to the equation (5), i.e. $p_r=m\rho$, the pressure
will vary linearly with the density and the range of the pressure will
differ from density with just by a factor of radial equation state parameter.
It is then clear from the Table 1 along with the Figs. (1) and (2)
that our bound is justified.

\subsection{TOV Equation}
For an anisotropic fluid distribution, the generalized TOV
equation is given by
\begin{equation}
\label{eq24} \frac{d}{dr}\left(p_r -\frac{\Lambda_r}{8\pi} \right)
+\frac{1}{2} \nu^\prime\left(\rho +p_r\right) +
\frac{2}{r}\left(p_r - p_t\right) = 0.
\end{equation}

Following Ponce de Le\'{o}n \cite{Leon1993}, we write the above
TOV equation as
\begin{eqnarray}
-\frac{M_G\left(\rho+p_r\right)}{r^2}e^{\frac{\lambda-\nu}{2}}-\frac{d}{dr}
\left(p_r -\frac{\Lambda_r}{8\pi} \right)
+\frac{2}{r}\left(p_t-p_r\right)=0, \label{eq25}
\end{eqnarray}
where $M_G=M_G(r)$ is the gravitational mass inside a
sphere of radius $r$ and is given by
\begin{equation}
M_G(r)=\frac{1}{2}r^2e^{\frac{\nu-\lambda}{2}}\nu^{\prime},\label{eq26}
\end{equation}
which can easily be derived from the Tolman-Whittaker formula and
the Einstein field equations. Obviously, the modified TOV
equation describes the equilibrium condition for the compact star
subject to the gravitational and hydrostatic plus another force due to
the anisotropic nature of the stellar object. Now, the above
equation can be written as
\begin{equation}
 F_g+ F_h + F_a=0,\label{eq27}
\end{equation}
where
\begin{eqnarray}
F_g &=& -B r\left(\rho+p_r\right), \label{eq28}\\
F_h &=& -\frac{d}{dr}\left(p_r -\frac{\Lambda_r}{8\pi} \right) \nonumber\\
&=& \frac{re^{-Ar^2}}{4\pi (m+1)}\left[ (m+1)\left(\frac{1}{Ar^4}+\frac{1}{r^2}\right) \right.  \nonumber\\
& &\left. + 2B(mA+1)+2mA(A-1) \frac{}{} \right]  - \frac{1}{4\pi r^3}, \label{eq29}\\
F_a &=& \frac{2}{r}\left(p_t -p_r\right). \label{eq30}
\end{eqnarray}

Note that here the variable $\Lambda$ contributes to the
hydrostatic force. The profiles of $F_g$, $F_h$ and $F_a$ for our
chosen source are shown in Fig. 5. The figure indicates that the static
equilibrium is attainable due to the combined effect of
pressure anisotropy, gravitational and hydrostatic forces.

\begin{figure}
\centering
\includegraphics[scale=.3]{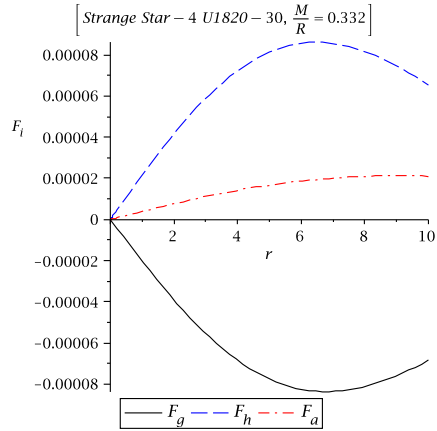}
\caption{Behaviours of pressure anisotropy, gravitational and hydrostatic
forces at the stellar interior of Strange star candidate $4U~1820-30$}
\label{fig:5}
\end{figure}

\subsection{Stability}
 In our anisotropic model, we define sound speeds as
\begin{equation}
\label{eq31} v_{sr}^{2} ~=~ m,
\end{equation}

\begin{eqnarray}
v^2_{st}=\frac{dp_t}{d\rho} =\frac{(m+1)}{2A(A+B)}\left[
\frac{e^{Ar^2}}{r^4}+A(B^2-AB)r^2\right. \nonumber\\
\left. - ((B^2-AB)+\frac{A^2-Am(2B+A)}{(m+1)})
 -\frac{A}{r^2}-\frac{1}{r^4}\right]. \label{eq32}
\end{eqnarray}

We plot the radial and transverse sound speeds in the Fig. 6 and
observe that these parameters satisfy the inequalities $0\leq
v_{sr}^2 \leq 1$ and $0\leq v_{st}^2 \leq 1$ everywhere within the
stellar object which obeys the anisotropic fluid
models \cite{Herrera1992,Abreu2007}.

The above equations (\ref{eq31}) and (\ref{eq32}) therefore lead to
\begin{eqnarray}
\label{eq33} v^2_{st}-v^2_{sr}=\frac{(m+1)}{2A(A+B)}\left[
\frac{e^{Ar^2}}{r^4}+A(B^2-AB)r^2\right. \nonumber\\
\left. -\left((B^2-AB)+\frac{A^2-Am(2B+A)}{(m+1)}\right)-\frac{A}{r^2}-\frac{1}{r^4}\right]-m
\end{eqnarray}

From the Fig. (7), we can easily say that $v^2_{st}-v^2_{sr}\leq 1$.
Since, $0\leq v_{sr}^2 \leq 1$ and $0\leq v_{st}^2 \leq 1$,
therefore,  $\mid v_{st}^2 - v_{sr}^2 \mid \leq 1 $.

Again, to check whether local anisotropic matter distribution is
stable or not, we use the proposal of \cite{Herrera1992}, known
as cracking (or overturning) concept, which states that the
potentially stable region is that region where radial speed of
sound is greater than the transverse speed of sound. In our case,
Figs. (6) and (7) indicate that there is no change of sign for the
term $v_{st}^2 - v_{sr}^2 $ within the specific configuration.
Therefore, we conclude that our compact star model is stable.

\begin{figure}
\centering
\includegraphics[scale=.3]{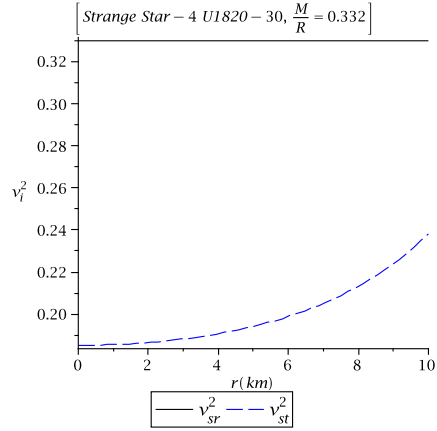}
\caption{Variation of radial and transverse sound speed of
Strange star candidate $4U~1820-30$}
\label{fig:6}
\end{figure}

\begin{figure}
\centering
\includegraphics[scale=.3]{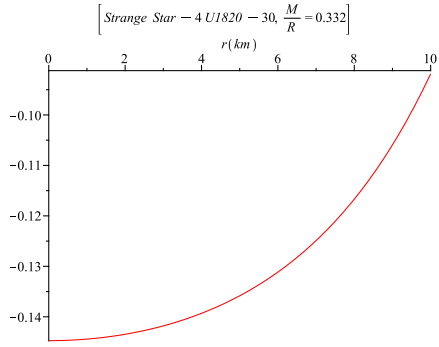}
\caption{Variation of $v_{st}^2-v_{sr}^2$ of Strange star candidate $4U~1820-30$}
\label{fig:7}
\end{figure}

\begin{figure}
\centering
\includegraphics[scale=.3]{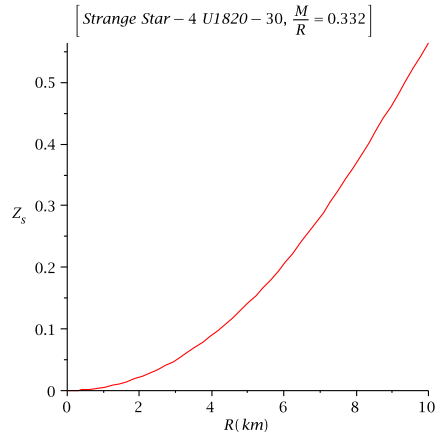}
\caption{ Variation of surface redshift of Strange star candidate $4U~1820-30$} \label{fig:8}
\end{figure}

\subsection{Surface Redshift}
The compactness of the star is given by
\begin{equation}
\label{eq35} u= \frac{M} {b}= - \frac{(A+B)}{4b(m+1)A^{\frac{5}{2}}}\left[2be^{-Ab^2}A^{\frac{3}{2}}+
\sqrt{\pi}~ erf(\sqrt{A}b)A\right].
\end{equation}
The surface redshift ($Z_s$) corresponding to the above
compactness ($u$) is obtained as
\begin{equation}
 1+Z_s= [1-2u]^{-\frac{1}{2}}, \label{eq36}
\end{equation}
where,
\[1+Z_s =
\left[1+\frac{(A+B)}{2b(m+1)A^{\frac{5}{2}}}\left(2be^{-Ab^2}A^{\frac{3}{2}}+
\sqrt{\pi}~ erf(\sqrt{A}b)A\right)\right]^{-\frac{1}{2}}.\]

\begin{eqnarray} \label{eq37} \end{eqnarray}

Thus, the maximum surface redshift for a Compact Star $4U~1820-30$
of radius $10$~km turns out to be $Z_s = 0.022$ (see Fig. 8).

\section{Concluding Remarks}

 In the present work we have performed some investigations regarding
the nature of compact stars after considering the following three inputs:

(1) The stars are anisotropic in their configurations, so that
$p_r \neq p_t$.

(2)  The erstwhile cosmological  constant, introduced by Einstein,
is variable in it's character, such that $\Lambda ~=~ \Lambda(r)
~=~\Lambda_r$.
This assumption of variable $\Lambda$ is the outcome of the
present day accelerating Universe in the form of so-called dark energy.

(3) The space-time of the interior of the compact stars can be described by KB metric.

In the investigations we have observed some interesting points which are as follows:

(i) Like normal matter distribution, the bound on the effective
EOS in this construction can be given by $0< \omega_i(r)<1$,
despite of the fact that star is constituted by the combination of
ordinary matter and effect of $\Lambda$.

(ii) As one of the dark enegry candidates the cosmological
variable $\Lambda$ contributes to the hydrostatic force for
stability of the compact stars.   However, in this connection we
realize that from the point of view of the physical interpretation
it would be better to consider the dark energy as being
represented by a scalar field, which would lead to the
interpretation of $\Lambda(r)$ as (also) being an effective
description of a bosonic condensate inside the star. Literature
survey shows that Boson stars or Bose-Einstein condensate stars
have been intensively investigated by several investigators
\cite{Chavanis2011,Hartmanna2012,Liebling2012}. So, there is a
scope to find out the possible connections between our model and
the description of Boson stars. Condensates as discussed by
Chavanis and Harko \cite{Chavanis2011} have maximum masses of the
order of $2~M_{sun}$, maximum central density of the order of
$(0.1-0.3) x 10^{16}~g/cm^3$ and minimum radii in the range of
$10-20$~km are comparable to the parameters in our model-based
some of the candidates of Strange Quark Star as shown in the
Table-1.

(iii) By applying the cracking concept of \cite{Herrera1992}
stability of the model has been attained surprisingly.

(iv)  The surface redshift analysis for our case shows that for
the compact star like $4U~1820-30$ of radius $10$ km turns out to
be $0.022$. In the isotropic case and in the absence of the
cosmological constant it has been shown that $z \leq 2$
\cite{Buchdahl1959,Straumann1984,Boehmer2006}. B{\"o}hmer and
Harko \cite{Boehmer2006} argued that for an anisotropic star in
the presence of a cosmological constant the surface red shift must
obey the general restriction $z \leq 5$, which is consistent with
the bound $z \leq 5.211$ as obtained by Ivanov~\cite{Ivanov2002}.
Therefore, for an anisotropic star with cosmological constant the
value $z = 0.022$ seems to be too low and needs further
investigation.

(v) The normal pressure $p_r$ vanishes but tangential pressure
$p_t$ does not vanish  at the boundary $r = b$ (radius of the
star). However, the normal pressure is equal to the tangential
pressure at the centre of the fluid sphere.

The overall observation is that our proposed model satisfies all physical requirements as well
as horizon-free and stable. The entire analysis has been performed in connection to direct
comparison of some of the Strange/Quark star candidates, e.g.
$Her~X-1$, $SAX~J~1808.4-3658(SS1)$ and $4U~1820-30$ which confirms validity of the
present model. Thus our approach lead to a better
analytical description of the Strange/Quark stars.

\section*{Acknowledgments} MK, FR, SR and SMH gratefully acknowledge support
from IUCAA, Pune, India under which a part of this work was carried out.
FR is also thankful to PURSE and UGC for providing financial support. We are
expressing our deep gratitude to the anonymous referee for suggesting some
pertinent issues that have led to significant improvements of the manuscript.

\end{document}